\documentclass[%
 reprint,
superscriptaddress,
 amsmath,amssymb,
 aps,
floatfix,
]{revtex4-1}

\usepackage{graphicx}% Include figure files
\usepackage{dcolumn}% Align table columns on decimal point
\usepackage{bm}% bold math
 \usepackage[english]{babel}
\begin{document}

\preprint{APS/123-QED}

\title{Continuous 40 GW/cm$^2$ laser intensity in a near-concentric optical cavity}

\author{O. Schwartz}
\email{osip@berkeley.edu}
\author{J. J. Axelrod}
\author{P. Haslinger}
\affiliation{Physics Department, University of California, Berkeley}
\author{C. Ophus}
\affiliation{National Center for Electron Microscopy, Molecular Foundry, Lawrence Berkeley National Laboratory, Berkeley, California}
\author{R. M. Glaeser}
\affiliation{Lawrence Berkeley National Laboratory, Berkeley, California}
\affiliation{Molecular and Cell Biology Department, University of California, Berkeley}
\author{H. M{\"u}ller}
\affiliation{Physics Department, University of California, Berkeley}

\date{\today}
\begin{abstract}
	Manipulating free-space electron wave functions with laser fields can bring about new electron-optical elements for transmission electron microscopy. In particular, a Zernike phase plate would enable high-contrast imaging of soft matter, leading to new opportunities in structural biology and materials science. 
	A Zernike plate can be implemented using a tight, intense continuous laser focus that shifts the phase of the electron wave by the ponderomotive potential. Here, we use a near-concentric cavity to focus 7.5\,kW of circulating laser power at 1064\,nm into a $7\,\mu$m waist, setting a record for continuous wave laser intensity and establishing a pathway to ponderomotive phase contrast TEM. 
\end{abstract}

\pacs{Valid PACS appear here}% PACS, the Physics and Astronomy
% Classification Scheme.
%\keywords{Suggested keywords}%Use showkeys class option if keyword
%display desired
\maketitle

%\tableofcontents
\section{Introduction}
Transmission electron microscopy (TEM) has emerged as a crucial source of  structural information with atomic resolution, both in molecular biology\cite{nogales_development_2016,glaeser_how_2016,callaway_revolution_2015} and  materials science\cite{urban_studying_2008,spence_high-resolution_2013}. One limitation of TEM is that specimens consisting of light elements, such as biological macromolecules, are nearly transparent to the electron beam, leading to weak image contrast. 
In optical microscopy, the problem of observing thin transparent objects, such as living cells, was solved by the invention of phase contrast microscopy by Zernike\cite{zernike_phase_1942}.   Introducing Zernike-type phase contrast to electron microscopy has been a goal of an increasingly intense  research effort\cite{glaeser_invited_2013}. Recently, phase contrast in TEM has been spectacularly demonstrated with carbon foil-based phase plates\cite{asano_molecular_2015,danev_volta_2014,dai_visualizing_2013}. Nevertheless, there is still significant potential for improvement: exposure to the electron beam changes the properties of the carbon  foil over time, varying the contrast transfer function and limiting the time a phase plate can be optimally used for imaging.

Controlling free-space electron propagation with lasers\cite{echternkamp_ramsey-type_2016,jones_laser_2016} offers an alternative approach to electron optics. A charged particle traversing an intense laser field experiences small-scale oscillatory motion, resulting in an effective `ponderomotive' potential. Experiments with electron scattering on a standing laser wave have shown that the ponderomotive potential can be used to create a diffraction grating\cite{freimund_observation_2001} and a beam splitter\cite{freimund_bragg_2002} for electron beams. 
It has been proposed recently that a laser beam focused in the back focal plane of a TEM objective lens can serve as a Zernike phase plate \cite{muller_design_2010}.
Unlike material phase plates, a laser phase plate is inherently immune to charging and electron beam damage, and offers negligible electron loss. The possibility of rapidly changing the phase delay by varying the laser power is an additional advantage. 

Free-space manipulation of energetic electrons used in TEM requires very high laser intensity. The phase delay induced by a focused Gaussian laser beam can be calculated as 
\begin{equation}
	\label{eq:phase}
	\phi = \sqrt{8\pi}\;\frac{\alpha}{\beta \gamma}\;\frac{P}{m c\, \omega^2\, w}\;,
\end{equation}
where $\alpha$ is the fine structure constant, $c$ is the speed of light, $m$ is the electron mass,  $\beta$ and $\gamma$ are the electron's relativistic factors, $P$ is the beam power, $\omega$ is the laser angular frequency, and  $w$ is the beam waist. Another requirement for a ponderomotive phase plate is that the focal spot size should not exceed a few micrometers\cite{muller_design_2010}. According to eq.\,(\ref{eq:phase}), imparting a $\frac{\pi}{2}$ phase shift to electrons at a typical TEM energy of 200-300 keV over a distance of several microns necessitates a laser intensity in the range of a few hundred GW/cm$^2$. Consequently, most experiments with electron scattering on light have been conducted with pulsed laser systems\cite{batelaan_illuminating_2007}. However, continuous operation is desirable for cryo-EM and other high resolution TEM applications where the signal to noise ratio is the limiting factor\cite{park_3d_2015,huang_sub-angstrom-resolution_2009}.

The laser power of a continuous-wave (CW) system can be enhanced using a power build-up cavity. Low-loss cavity mirrors have been shown to withstand intensities up to 0.1\,GW/cm$^2$ \cite{meng_damage_2005}. Much higher intensities required for a ponderomotive phase plate can be achieved in a focusing cavity, such as a near-concentric Fabry-P\'{e}rot resonator. In this configuration, the fundamental mode has an hourglass shape, with laser power concentrated in a small focal spot at the center but spread out over a large area on the mirror surface, which prevents mirror damage. Tight intra-cavity focusing at low power has been demonstrated in a medium-finesse near-concentric cavity\cite{durak_diffraction-limited_2014}. At the same time,  average circulating power of up to 670\,kW\cite{carstens_megawatt-scale_2014} has been achieved in a focusing cavity built for amplifying trains of ultrashort pulses for intra-cavity high harmonic generation and optical comb spectroscopy in the extreme ultra-violet spectral range \cite{cingoz_direct_2012}. However, a combination of high power and tight intra-cavity focusing necessary for a laser phase plate has not been realized yet.

\begin{figure}
	%\hspace{1in}
	\centering
	\includegraphics[width = 3.5in]{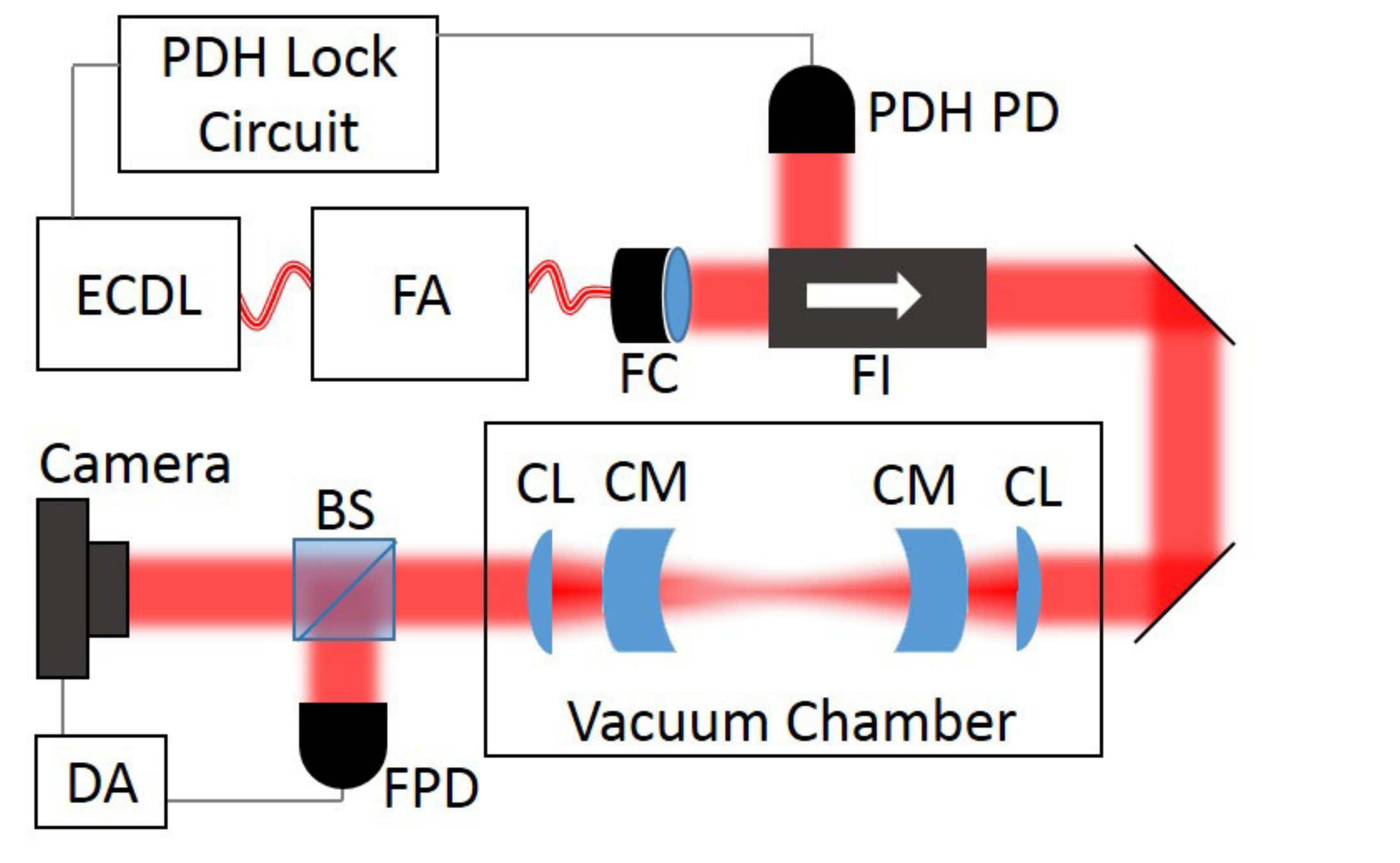} 
	\caption{Schematic diagram of the experimental setup. ECDL = external cavity diode laser, FA = fiber amplifier, FC = fiber coupler, PDH PD = Pound-Drever-Hall lock photodiode, FI = Faraday isolator, CL = coupling lens, CM = cavity mirror, BS = beamsplitter, FPD = fast photodiode, DA = digital data acquisition.}
	\label{fig:1}
\end{figure}

Here, we report reaching a milestone towards a prototype laser phase plate, implemented as a high finesse,  high numerical aperture near-concentric cavity.  We characterize its fundamental mode and use a numerical model to analyze properties of TEM in the presence of the intra-cavity laser field. With 7.5 kW of circulating CW laser power, we demonstrate a maximum intensity of 41\,GW/cm$^2$, previously achieved only in pulsed laser systems, which is sufficient to retard a 300 keV electron beam by 0.16\,rad. 

\section{Experimental Results}
Our optical system, shown schematically in Fig.\,\ref{fig:1}, consists of a near-concentric cavity and a CW feeding laser, operating at a wavelength $\lambda=1064$\,nm. The feeding laser comprises an external cavity diode laser, frequency-locked to the cavity using the Pound-Drever-Hall method, and a fiber amplifier. The cavity is designed for insertion into a plane conjugate to the back focal plane of a TEM objective lens, with the electron beam entering the cavity orthogonally to the optical axis. The cavity is formed by two concave mirrors (Layertec) with a diameter of $12.7$\,mm,  radius of curvature  of $12.7$\,mm, and specified reflectivity  $\mathcal{R}=1 - (10\pm 5)\cdot 10^{-5}$. The back surface of the mirrors is convex, concentric with the front surface. The meniscus shape allows for efficient coupling into the high numerical-aperture mode of the cavity with a single aspheric lens. 

The cavity mount allows for adjustment of the tilt and axial position of one of the mirrors, housed in a flexure suspension. The entire cavity housing is machined from a single block of aluminum, to ensure precise centering of the mirrors and to provide effective thermal conduction to cool the cavity. Alignment of the near-concentric cavity, requiring angular precision better than 1\,$\mu$rad, is achieved by three fine-pitched micrometer screws providing rough alignment, pressing against three piezo actuators positioned in the pockets of the aluminum block. The high-power optical module comprising the cavity, the coupling lenses and the mirror alignment optomechanics, is made compact enough to fit into a cylindrical space of $25$\,mm diameter, which facilitates future integration into a TEM system.

\begin{figure}
	%\hspace{1in}
	\centering
	\includegraphics[width = 3.5in]{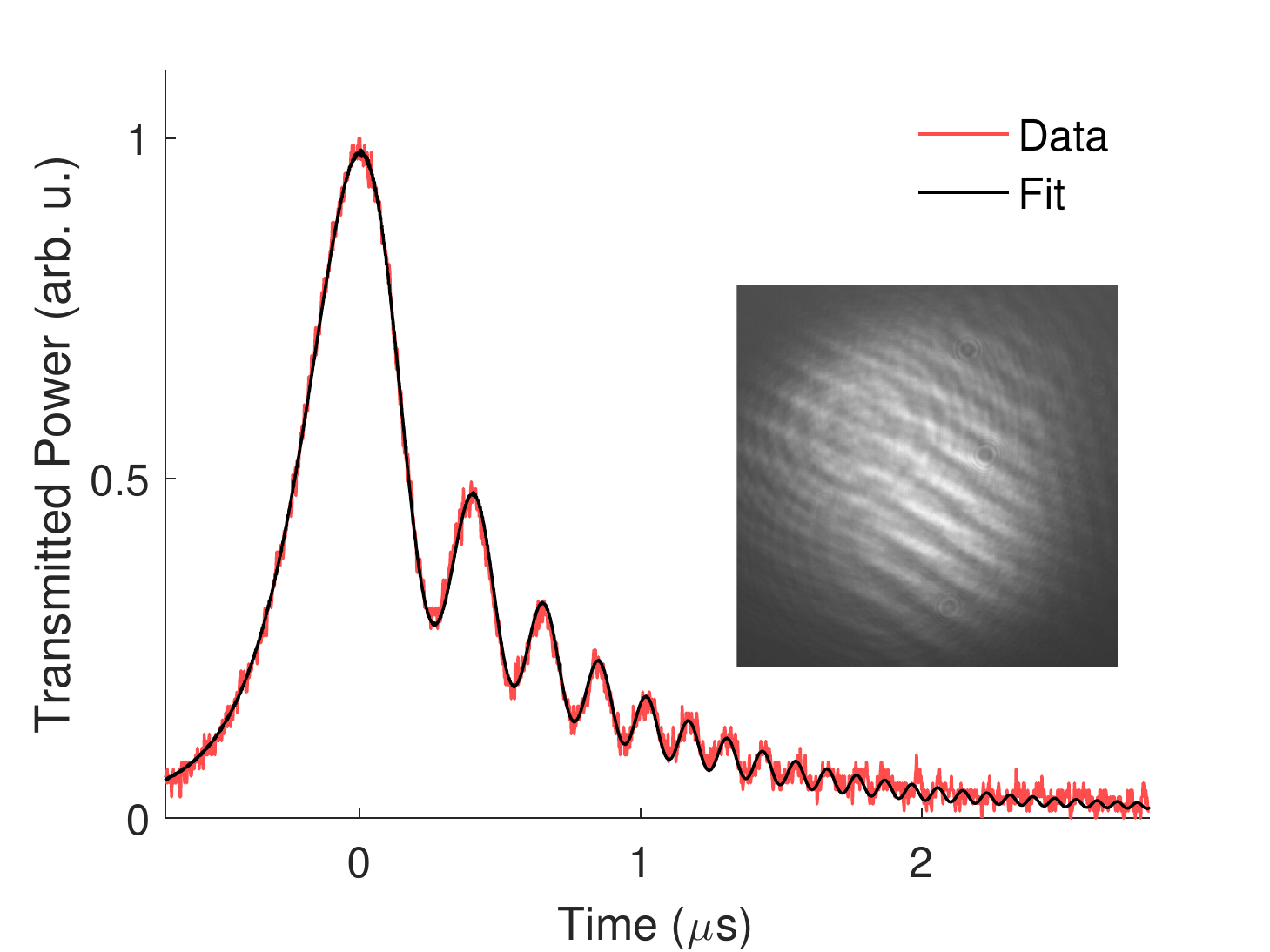} 
	\caption{\textbf{Axes:} The cavity ring-down profile observed at $\mbox{NA}_a=0.0469$, $\mbox{NA}_b = 0.0524$ (red), and fit to the model described by eq. \eqref{e1} (black). The fit corresponds to a cavity mirror transmission-plus-loss of $137.9 \pm 0.4\,\mbox{ppm}$. \textbf{Inset:} The mode for which the displayed CRD profile was measured. Fringes are an imaging artifact due to interference from the camera's detector window.}
	\label{fig:2}
\end{figure}

The cavity is suspended in a vacuum chamber pumped down to $2\cdot 10^{-7}$\,mbar, emulating the environment of an existing TEM column and preventing undesirable ionization of air molecules.
Using the tilt and axial motion degrees of freedom of one of the mirrors, the cavity was brought to a near-concentric configuration. To characterize the size of the focal point inside the cavity, we tuned the laser frequency to oscillate around the fundamental mode of the cavity. The transmitted beam was collimated by an  aspheric lens (focal length 25\,mm) and directed into a CMOS image sensor. Fitting the mode image with a two-dimensional Gaussian profile, we obtain the width of the fundamental mode at the far field, reciprocal to the size of the focal spot. The image, shown in Fig. \ref{fig:2}, exhibits a small degree of ellipticity determined by a very slight astigmatism of the cavity mirrors. The two principal axes of the ellipse correspond to numerical apertures of $\mbox{NA}_a = 0.0469 \pm 0.0005$ and $\mbox{NA}_b=0.0524 \pm 0.0005$. The mode waist corresponding to $\mbox{NA}_b$ is $s=\lambda \left(\pi \text{\small{NA}}\right)^{-1} =6.46\,\mu$m.

The reflectivity of the cavity mirrors was measured using the cavity ring-down (CRD) method \cite{berden_cavity_2000}, in which light is briefly injected into the cavity and the subsequent rise and decay in the power of the transmitted light is observed. To avoid the need for a pulsed laser source or optical modulators, injection of light into the cavity was accomplished by rapidly sweeping the laser frequency across a longitudinal mode resonance of the cavity's $\mbox{TEM}_{00}$ mode (rapidly-swept cw-CRD) \cite{orr_rapidly_2011}. 

Under these conditions, the transmitted electric field amplitude is well-modeled by the inverse Fourier transform of the product of the cavity's transfer function and the spectrum of a linearly-chirped laser field, so that the transmitted power is
\begin{equation}
	P \left( t\right) \propto \left| \int_{-\infty}^{\infty} d\omega \, e^{i \omega t} \cdot \left[  \frac{ e^{-i \omega L/c} }{1 - \mathcal{R} e^{- 2 i \omega L /c}  }    \right] \cdot \left[ e^{ - i \omega^2/2\eta }  \right] \right|^2 \label{e1}
\end{equation}
where $\mathcal{R}$ is the cavity mirror reflectance, $L$ is the cavity length, and $\eta$ is the frequency sweep rate. This model was used to fit the experimentally measured CRD profiles, with $\mathcal{R}$ serving as the fit parameter of interest.

The measured CRD profile is shown in Fig. \ref{fig:2} along with its least-squares best fit to the model described by \eqref{e1}. Expressing the cavity mirror reflectivity in terms of the cavity mirror transmission, $T$, and loss, $L$, such that $\mathcal{R}=1-(T+L)$, the fitted profile corresponds to a cavity mirror transmission-plus-loss of $137.9 \pm 0.4\,\mbox{ppm}$. This corresponds to a cavity finesse of $\mathcal{F} = \frac{\pi \mathcal{R}}{1-\mathcal{R}} \approx \frac{\pi}{T+L} = 22780 \pm 65$.

The seed laser was locked to the cavity using the Pound-Drever-Hall method, with side bands generated by direct RF modulation of the seed laser current. The reflected beam was separated by a Faraday isolator and directed into a photodiode. The RF signal from the diode was demodulated and used as the error signal. 

\begin{figure}
	\centering
	\includegraphics[width = 3.5in]{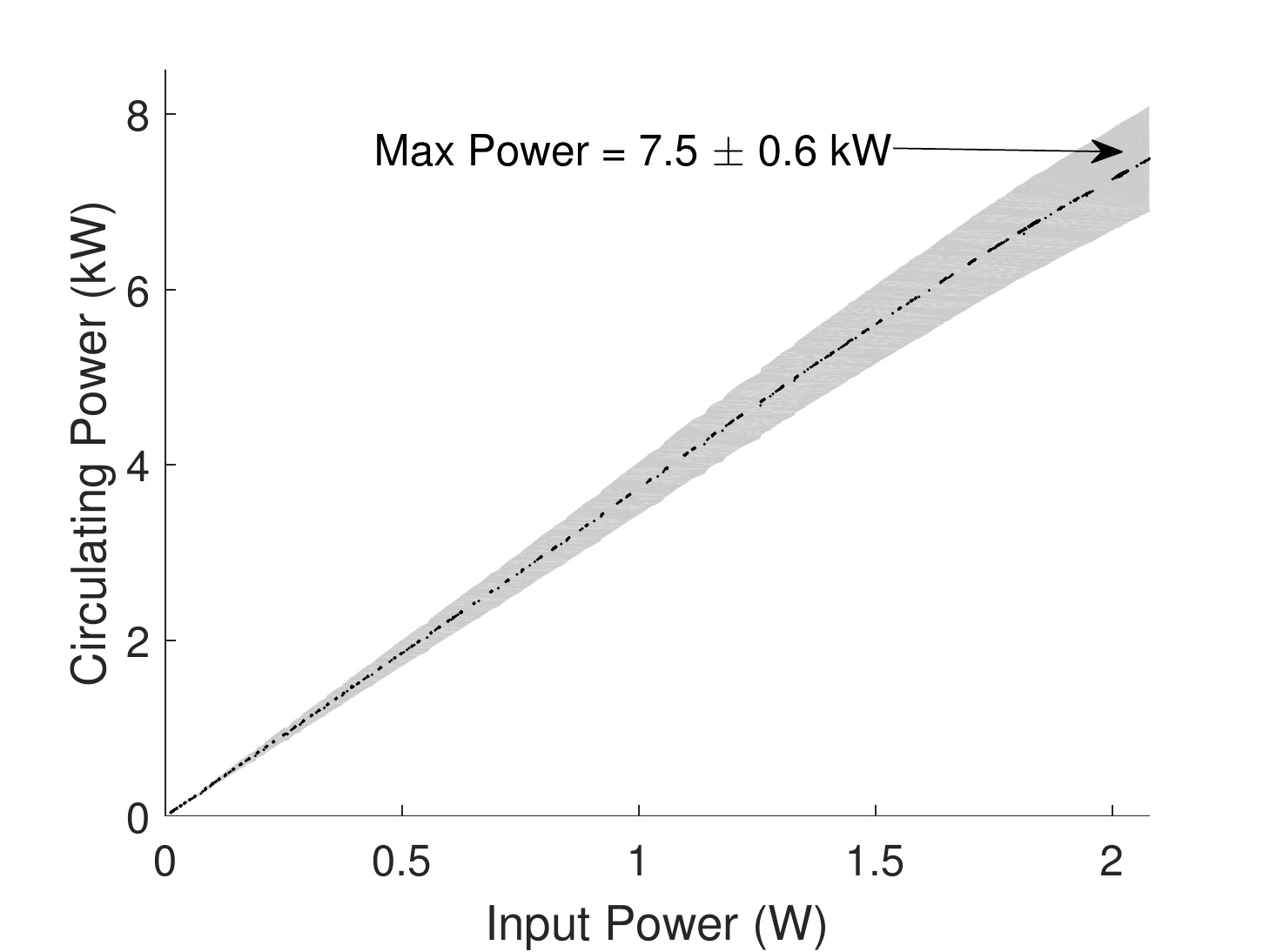} 
	\caption{Intra-cavity circulating power as a function of input power. The gray region represents the 68\% confidence interval. The circulating power is determined via the measurement of the amplification factor $M = \left( T_{cav} + 1 - R_{cav}\right)/ 2(T+L)$. Here, $T_{cav}$ was monitored as a function of input power by simultaneously measuring the input power (using a calibrated partially-reflective window) and the transmitted power. $R_{cav}$ was measured once at low power.}
	\label{fig:3}
\end{figure}

To estimate the circulating power in the cavity, in addition to cavity finesse we need to know the coupling efficiency and the transmission-to-loss ratio of the mirrors. Both parameters can be inferred from measurements of the cavity transmission coefficient $T_{cav}$ and reflection coefficient $R_{cav}$. Denoting the mode overlap between the input beam and the fundamental cavity mode as $Q$, we have: 
\begin{equation}
	T_{cav} = \left|Q\right|^2\left(\frac{T}{T+L}\right)^2,\;\; R_{cav} =  1 - 2 \left|Q\right|^2\frac{T}{T+L}  + T_{cav}
\end{equation}
With the laser frequency locked to the cavity resonance, we measured $R_{cav}=0.34 \pm 0.03$, $T_{cav}=0.32 \pm 0.03$. Extracting the cavity parameters, we get $\left|Q\right|^2=0.75 \pm 0.05$ and $\frac{T}{T+L}=0.65 \pm 0.05$. 
Taken together with the CRD data, these parameters allow us to determine the mirror transmission $T=90 \pm 7\,$ppm and the amplification factor as $M= \left|Q\right|^2 \frac{T}{\left(T+L\right)^2} = \left( T_{cav} + 1 - R_{cav}\right) / (2 (T+L))= 3600 \pm 150$.

%\subsection{High power operation}
With the cavity parameters determined, we proceeded to increase the input power. With the cavity chamber held at atmospheric pressure, increasing the input power beyond 300\,mW did not lead to further increase in transmitted power, apparently due to the onset of nonlinear optical effects in air at a circulating power of about 1kW, corresponding to a maximum intensity of 5.5\,GW/cm$^2$. With the chamber evacuated, we were able to reach intra-cavity power of up to $7.5 \pm 0.6$ kW. The intra-cavity circulating power (inferred from transmitted power) as a function of input power is shown in Fig. \ref{fig:3}. The graph is almost linear, with a small deviation at higher power possibly arising due to thermally induced deformations of the cavity housing modifying the cavity alignment. The maximum power was limited by the concern over the risk of thermal damage to the mirrors, which were not well thermally coupled to the mount.
With the mode parameters measured above, the maximum measured power corresponds to a maximum intensity of $(41 \pm 4) \,\text{GW}/\text{cm}^2$, which would lead to a phase retardation of $0.16$\,rad for a 300\,kV electron beam. Repeated CRD measurements at low power confirmed that no damage to the mirrors occurred during the high power run.
\section{Numerical Modeling}
\begin{figure}
	\centering
	\includegraphics[width = \columnwidth]{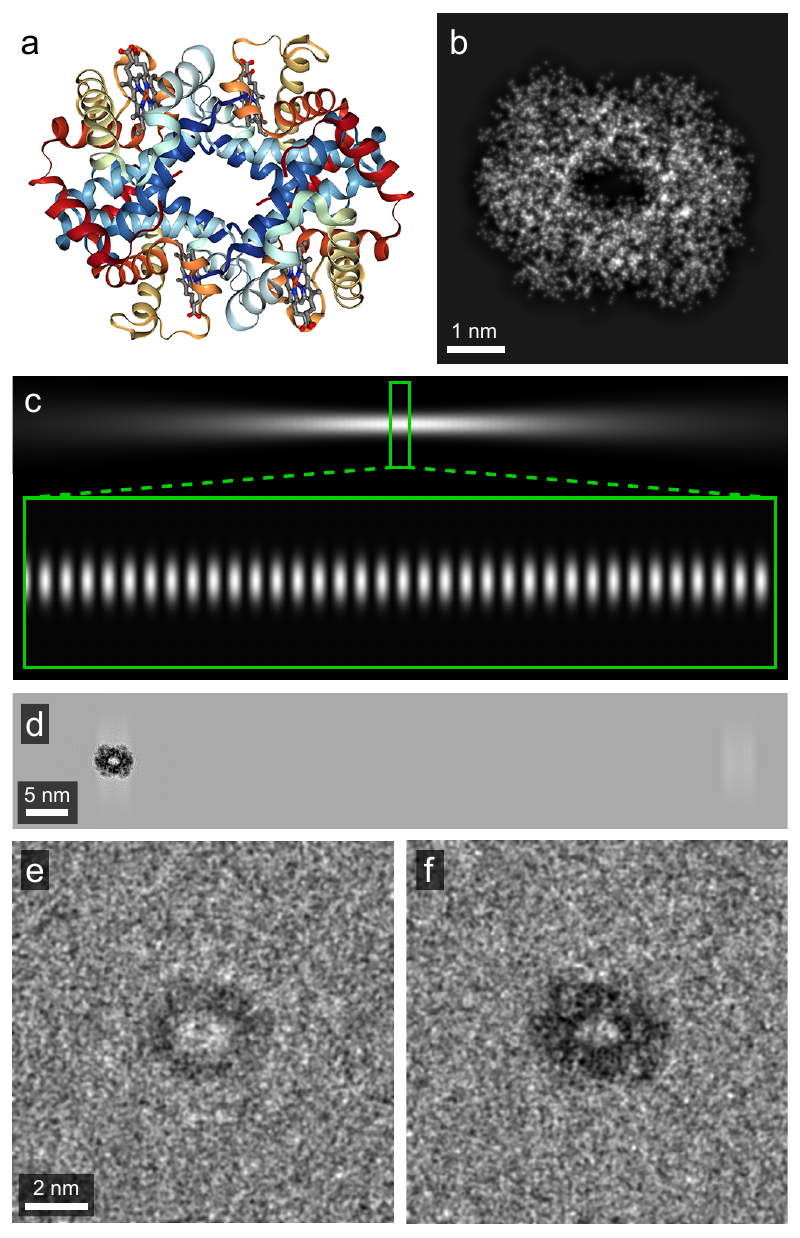} 
	\caption{Modeling of TEM images of a hemoglobin molecule with a cavity-based ponderomotive phase plate: (a) a ribbon diagram of the molecule, (b) a two-dimensional projection of the atomic potential, (c) the phase shift caused by the fundamental mode of the optical cavity. The magnified image shows individual fringes of the standing wave. Panel (d) shows a simulated TEM image, without the shot noise, of the molecule in the same orientation as in (a,b), with a shifted `ghost' image corresponding to the first diffraction order.  A simulated conventional TEM image, underfocused  by $1\,\mu$m, at a dose of 20$\text{e}/\text{\AA}^2$, is shown in (e).  Panel (f) shows an in-focus image formed with a cavity-based ponderomotive phase plate at the same electron dose. 
		The cavity numerical aperture in the model is $NA=0.05$, in agreement with the experimentally demonstrated parameters.  The model assumes that the intra-cavity power is scaled to achieve full $\frac{\pi}{2}$ retardation at maximum, which requires a roughly tenfold further increase of optical power. }
	\label{fig:4}
\end{figure}
To evaluate the effect of the laser phase plate on TEM of biological macromolecules, we have conducted numerical simulations of TEM imaging of human hemoglobin embedded in vitreous ice. This tetrameric complex has a molecular mass of approximately $64$\,kDa, which is too small for conventional TEM reconstruction, but has been recently solved to $3.2$\,\AA\ resolution using phase contrast TEM with a carbon foil phase plate\cite{khoshouei_cryo-em_2016}.

%Colin's text
Multislice simulations were performed using the methods and potentials described in Kirkland\cite{kirkland_advanced_2010}. An accelerating voltage of 300 kV, a pixel size of $0.2$\,\AA\ and spherical aberration of $1.3$\,mm was used.  Because shot noise and thermal smearing of the potentials dominated the information limit of the simulations, the finite spatial and temporal coherence of the electron beam were neglected in the simulations. The hemoglobin structure used was downloaded from the protein databank\cite{kavanaugh_accommodation_1993}. The thermal vibration of the protein atoms was assumed to be $0.1$\,\AA, applied as an envelope function. The continuum model of vitreous ice developed by Shang and Sigworth\cite{shang_hydration_2012} was used to model the embedding potential around the hemoglobin structure, numerically integrated in 3D. 
%End Colin's text

The results of the modeling are presented in Fig.\,\ref{fig:4}. The ribbon model of the ring-like protein complex and the projected potential map are shown in Fig.\,\ref{fig:4}\,(a,b). The spatial map of the phase shift induced by the Gaussian standing wave cavity mode is shown in Fig.\,\ref{fig:4}\,(c), with a zoom-in plot showing the individual minima and maxima of the standing wave.  Simulated images of hemoglobin, seen in the same projection as \ref{fig:4}\,(b) are shown in Fig.\,\ref{fig:4}(e,f). 

A side effect of passing an electron beam through a
standing laser wave is that the standing wave acts as a diffraction grating
for the electrons, which generates additional weak `ghost' images. These ghost images are displaced from the primary image by a distance  $\delta x = 2 n f \lambda_e/\lambda$, where $n$ is the diffraction order, $f$ is the focal distance of the TEM objective and $\lambda_e$ is the electron wavelength. A first order ghost image is shown on the right side of Fig.\,\ref{fig:4}(d). This panel is shown without the shot noise, which would otherwise render the ghost image nearly invisible. Since the amplitude of such ghost images is well below shot noise, they will not be visible in individual images, and will amount to an inconsequential contribution to the noise in the averaged images used for density map reconstruction. 

Panel (e) shows a simulated TEM image of the hemoglobin molecule. Highly transparent biological
macromolecules are conventionally made visible by defocusing
the imaging system from the specimen plane,
creating a phase contrast image with an oscillatory contrast
transfer function\cite{glaeser_electron_2007}. While a higher defocus  results in higher contrast at low spatial frequencies, it also leads to a loss of contrast at high spatial frequencies. The defocus of $1\,\mu$m used here is a value that still allows for reconstruction of density maps with near-atomic resolution\cite{cheng_single-particle_2015,merk_breaking_2016}. The shot noise is modeled assuming an effective dose of 20$\text{e}/\text{\AA}^2$,  typically used in TEM protein structure studies as an optimum point between radiation damage, which increases with the dose, and the shot noise decreasing with it. Fig.\,\ref{fig:4}(f) shows an in-focus image of hemoglobin with the ponderomotive phase plate at the same electron dose. A full $\pi/2$ phase shift at the intensity maximum is assumed. The phase contrast image demonstrates a stronger signal at low spatial frequencies compared to a defocus-contrast image, which is expected to enable particle projection classification and alignment for macromolecules at least as small as hemoglobin.

\section{Outlook}
The numerical results shown in Fig. \ref{fig:4} demonstrate that a standing wave built up in a focusing resonator creates a contrast transfer function suitable for phase contrast imaging. Importantly, the well-defined spatial structure of the cavity mode ensures that the contrast transfer function can be accurately taken into account when interpreting the EM images. While the intensity demonstrated in our experiment is about an order of magnitude below that required to impart a $\frac{\pi}{2}$ phase shift to a $300$\,keV electron beam, it may be sufficient for the initial demonstration of the ponderomotive retardation. Furthermore, using state of the art mirrors it should be possible to increase the cavity finesse to $2\!\cdot\! 10^5$. Increasing the input power to 30  W, which is possible with commercial fiber amplifiers at NIR wavelength, should be sufficient to increase the focal intensity to well beyond $10^{12}$\,W/cm$^2$. 

In this work, we have focused on developing a laser-based Zernike phase plate. However, a number of other tools can be envisioned using high-intensity intra-cavity CW laser fields. For example, a quantum imaging method based on an interaction-free measurement scheme has been proposed\cite{kruit_designs_2016}. A significant obstacle to implementing this scheme lies in the absence of a high quality beam splitter for the electron wave function. A CW standing wave  inside an optical cavity can act as a highly regular, virtually lossless phase grating, coherently splitting an electron beam into two paths via Kapitza-Dirac scattering in the Bragg regime\cite{freimund_bragg_2002}. Such a beam splitter could also enable various electron interferometry schemes, mimicking the diverse family of optical interferometers used for metrology and sensing.

Finally, we note that the type of cavity we have built can be of interest to a wide class of experiments. The combination of a small mode volume with the open, accessible geometry of a high-NA near-concentric resonator can be useful for cavity QED experiments\cite{wolke_cavity_2012,stute_tunable_2012}. Furthermore, the ability to build up very high circulating power can be used to implement ultra-deep dipole traps\cite{edmunds_deep_2013, chen_measurement_2015}, as well as for trapping and cooling  nanoparticles\cite{chang_cavity_2010, asenbaum_cavity_2013, kiesel_cavity_2013}.

In summary, we have developed a high-finesse optical cavity with a tightly focused fundamental mode. We verified by numeric simulations that such field configuration can function as a ponderomotive phase plate for TEM. We have demonstrated that optical intensity in the range of tens of GW/cm$^2$ can be reached in a CW laser system using a near-concentric Fabry-P\'{e}rot resonator.
These results represent a  significant step towards ponderomotive phase contrast TEM, and, more generally, pave the way towards laser based coherent control of free space electron wave functions.
\section*{Acknowledgments}
This material is based upon work supported by the National Science Foundation under Grant No. 1040543. OS was supported by a Human Frontier Science Program postdoctoral fellowship LT000844/2016-C. PH thanks the Austrian Science Fund (FWF): J3680. Work at the Molecular Foundry was supported by the Office of Science, Office of Basic Energy Sciences, of the U.S. Department of Energy under Contract No. DE-AC02-05CH11231.

\end{document}